# BIOMECHANICAL CARACTERISATION OF LUMBAR BELT BY FULL-FIELD TECHNIQUES: PRELIMINARY RESULTS

R. Bonnaire[a,b], J. Molimard[a], P. Calmels[c] and R. Convert[b]

a LCG, UMR 5146, École Nationale Supérieure des Mines, CIS-EMSE, CNRS, Saint-Étienne, France
b Thuasne, 27 rue de la Jomayère 42031 St-Etienne, cedex 2, France
c Physical medicine and rehabilitation CHU Bellevue, 25 boulevard Pasteur 42055 St-Etienne, France
bonnaire@emse.fr, molimard@emse.fr, paul.calmels@chu-st-etienne.fr, reynald.convert@thuasne.fr

**ABSTRACT:** In France, 50% of the population per year is suffering from low back pain. Lumbar belt are frequently proposed as a part of the treatment of this pathology. However mechanical ways of working of this medical device is not clearly understood, but abdominal pressure is often related. So an optical method was developed in this study to measure strain in lumbar belt and trunk interface and to derive a pressure estimation. Optical method consisted of coupling fringe projection and digital image correlation (DIC). Measurement has been carried out on the right side of a manikin wearing a lumbar belt. Average strain is 0.2 and average pressure is 1 kPa. Continuation of this study will be comparison of strain and pressure in different areas of lumbar belt (left side, front and back) and comparison of different lumbar belts. Results will be used in a finite elements model to determine lumbar belt impact in intern body. In long term, this kind of study will be done on human.

## 1. INTRODUCTION

In France, 50% of population suffers of low back pain every year [1]. An important part of health care cost and cessation of work is due to low back pain [1-2]. Many treatments are proposed without high clinical evidence. There is no consensual guideline provided to physicians. Treatments depend on patient, low back pain type, evolution and physician knowledge, believes and modalities. Lumbar belt are proposed to treat subacute low back pain [3-4]. Nevertheless the physiological effects of these medical devices are not clearly understood.

The objective of this study is to characterize mechanical behaviour of lumbar belt using full-field methods: fringe projection coupled to Digital image correlation.

## 2. METHODOLOGY

### 2.1. Full-field measurement

To determine shape of an object, fringes are projected on a plane (reference) and the studied object. Object presence creates a phase shift related to the object depth. This relation is described by the equation

$$\phi(x,y) = S(x,y) \times z(x,y) = \frac{2\pi \times \tan\theta(x,y)}{p(x,y)} z(x,y) \qquad (1)$$

with $z(x,y)$ object depth, $S$ the sensitivity of the optical set-up, related to the illumination to observation angle and the projected grid step. A local calibration is needed to obtain $S$. The phase $\Phi(x,y)$ is calculated by phase shifting method.

Digital image correlation (DIC) goal is to determine displacement field between reference image im0 described by grey level $f(x,y)$ and a deformed image im1 described by grey level $g(x,y)$ expressed by:

$$g(x,y) = T(f(x,y)) \qquad (2)$$

where T refers to the mechanical transformation. Parameters of the transformation ($\delta x$, $\delta y$) are obtained through the maximization of the correlation product (f * g) defined by [5]:

$$h(r,s) = (g*f)(r,s) = \frac{\int_{-\infty}^{+\infty}\int_{-\infty}^{+\infty} g(a,b) f(a-r,b-s)\,dadb}{\int_{-\infty}^{+\infty}\int_{-\infty}^{+\infty} g(a,b)\,dadb \times \int_{-\infty}^{+\infty}\int_{-\infty}^{+\infty} f(a-r,b-s)\,dadb} \qquad (3)$$

Here, the maximum of the correlation product (f * g) is obtained in Fourier space, and sub-pixel interpolation by a generalized spatial phase shifting.

Fringe projection method permits to determine the shape of the object (x,y,z) and digital image correlation gives a displacement on the object plane of the camera. The post-processing used is based first on the projection of displacement obtained by DIC on this shape, leading to the (u,v,w) displacement fields in a global landmark. Displacement field in a local landmark denoted ($u_t$, $v_t$, $w_n$) is obtained using shape derivation. Last, strain is obtained from the displacement vector gradient in the local landmark.

## 2.2. Experimental protocol

Experimental set-up consists of 1280×960 pixels resolution camera, fringe projector and manikin wearing Lombaskin® (Thuasne, Saint-Etienne, France) lumbar belt. Interest zone is 95×70 mm² on the right side of manikin. Fig. 5 illustrates camera and projector disposition (a) and manikin wearing lumbar belt (b).

Experimental study was done in two steps. First step consists of optical acquisition with the lumbar belt worn not stretched; second step of optical acquisition with tight lumbar belt worn by a mannequin i.e. closed and stretched. Displacement difference permits to evaluate lumbar belt strain when tighten.

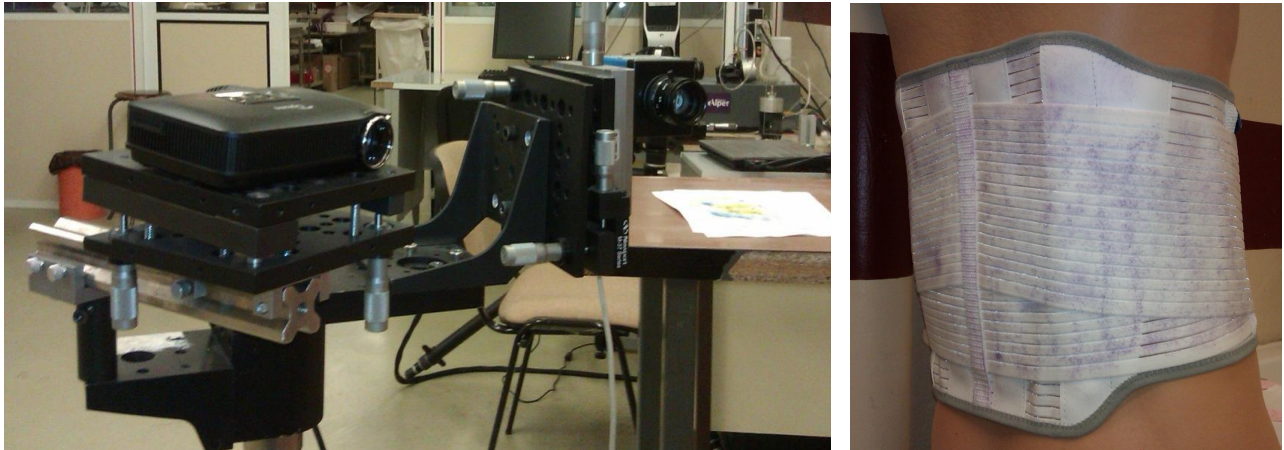

a/ b/

**Figure 1- a/ Camera and projector disposition, b/ Manikin wearing lumbar belt**

## 3. RESULTS AND DISCUSSION

Strain $E_{xx}$ is represented in Fig.2. Average of strain $E_{xx}$ is around 0.2. This value is close to the information given by the manufacturer in normal use of lumbar belt. Maximum strain $E_{xx}$ is found around whale, and its value is close to 1.

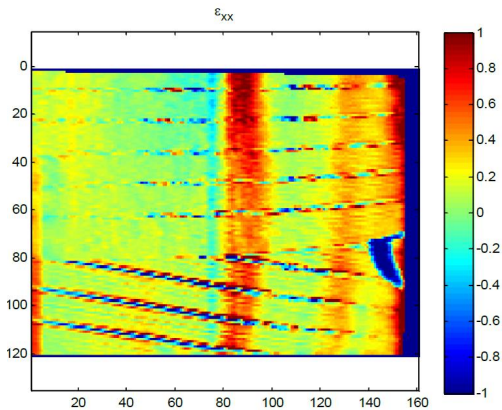
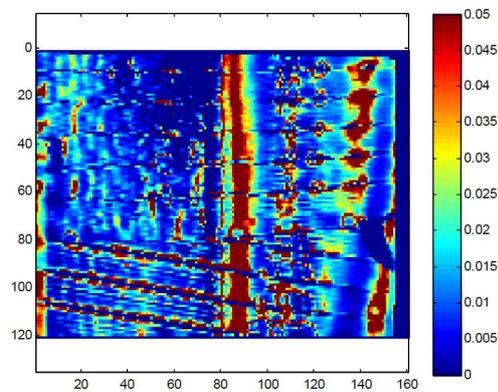

**Figure 2- Strain $E_{xx}$ in interested zone**  **Figure 3- Pressure in interested zone**

The pressure applied by the lumbar belt on the trunk is a key parameter to characterize its effect. For a given behaviour law, it is possible under a certain set of hypothesis to propose a first evaluation of the pressure field applied by the lumbar belt on the trunk by using the Laplace law [6]. It relates the pressure $P$, the linear tension $T$ and the curvature $R$ by the expression:

$$P = \frac{T}{R} \quad (4)$$

Curvature is obtained by double derivation and a second order polynomial identification. Pressure is represented in Fig.3. Mean pressure is around 1 kPa. This value is quite the same as a preliminary study of pressure measurement, but it is is worth noting that the pressure distribution is very heterogeneous, with pressure-free domains, and pressure peaks in the whale area.

## 4. CONCLUSION

This study is a preliminary approach. Average values (strain and pressure) when lumbar belt is tightening are in good agreement with the manufacturer's data, but the heterogeneity found in the strain and pressure fields indicates that a single value is insufficient to describe the lumbar belt behavior. The complete strain /pressure distributions are supposed to have a deterministic effect on the postural modification during daily movements, on the tolerance of the treatment and on other clinical aspects such as proprioceptive or antalgic effects. In other words, the lumbar belts are supposed to act in a therapeutically point of view on the back pain by modifying the posture, with a sufficient acceptability, i.e. a good level of comfort.

Last, the link between belt architecture, material properties, induced strain when tightening and applied pressure are of fundamental interest in the lumbar belt design and wearing. In the near future, the coupling of optical methods (for shape and strain estimation) and pressure matrices sensor must be done.

These results will be used in a 3D finite elements modeling to characterize lumbar belt impact on back physiological movement and constraints.